# Integrated terahertz photonic receiving frontend with link noise outperforming electronics

Yuansong Zeng[1,†], Zixi Wang[1,†], Liga Bai[2,†], Yuansheng Tao[1], Yiwen Zhang[1], Yifan Wu[1], Zhe Ding[2], Xiangzhi Xie[1], Zhaoxi Chen[1], Lu Zhang[2], Shi-Wei Qu[3], Jun Hu[3], Hanke Feng[1], Chi Hou Chan[1], Xianbin Yu[2,*], and Cheng Wang[1,*]

[1] *Department of Electrical Engineering & State Key Laboratory of Terahertz and Millimeter Waves, City University of Hong Kong, Kowloon, Hong Kong, China*

[2] *College of Information Science and Electronic Engineering, Zhejiang University, Hangzhou, China*

[3] *School of Electronic Science and Engineering, University of Electronic Science and Technology of China (UESTC), Chengdu, China.*

[†] *The authors contributed equally to this work*

**Corresponding author e-mail address: xyu@zju.edu.cn; cwang257@cityu.edu.hk*

## Abstract

Terahertz technology is a key enabler for sixth-generation (6G) wireless networks, yet its application is constrained by increasingly severe free-space loss at high frequencies. To efficiently retrieve weak signals at the receiving end, a compact frontend that features both a high-gain antenna and a low-noise signal-detection chain is critical. Current transistor-based THz electronic frontends face significant challenges in meeting these demands because both on-chip antenna efficiency and transistor noise performance degrade rapidly when approaching their cut-off frequencies. Photonic technology provides an alternative solution to circumvent the transistor bandwidth limit, yet most microwave photonic links to date exhibit noise performance substantially worse than state-of-the-art electronics. Here, we demonstrate low-noise integrated THz photonic frontends that deliver undegraded link noise performance across three major THz windows from 140 to 450 GHz, and outperform electronic frontends in the upper two windows. We achieve this through co-design of high-gain on-chip THz antenna array and broadband THz-optic modulator on a single thin-film lithium niobate (TFLN) chip, leading to distributed reception of free-space THz signals and continuous coherent build-up of the THz-optic conversion process with unprecedented efficiency. Combined with an efficient heterodyne detection chain, our integrated frontends exhibit effective isotropic noise figures of 13.6 and 16.2 dB at 250 and 450 GHz, respectively, both setting new benchmarks in their respective bands. We further demonstrate 6G-oriented multi-link communication up to 20 Git/s. Our integrated frontends represent a significant step towards compact, cost-effective and energy-efficient THz wireless systems in 6G and beyond.

## Introduction

The evolution towards sixth-generation (6G) wireless network is expected to move beyond the traditional single-lane race on peak data rate, toward ubiquitous connectivity with native support for integrated sensing and communications (ISAC)[1-3], enabling transformative applications from autonomous driving to intelligent manufacturing. Terahertz (THz) waves, with the potential to offer terabit data rate[4-6] and millimeter-level sensing resolution simultaneously[7,8], have become a key enabler to fulfill this vision[9-11]. However, THz waves suffer from severe free-space path loss and atmosphere absorption[12] that leave ultra-weak signals at the receiving end. To effectively retrieve these weak free-space signals, it is crucial to design receiver frontends that incorporate both a high-gain antenna to counteract free-space loss and a high-fidelity detection link for low-noise signal recovery. Furthermore, it is imperative to integrate both modules on a single chip-scale platform, which not only minimizes device-interconnection loss but also ensures compliance with the stringent requirements for compactness and cost-effectiveness in the 6G era. However, simple scaling of existing microwave/millimeter-wave circuitry into the THz regime faces significant challenges in meeting the above requirements due to limitations in both transistor noise and antenna gain. A typical electronic signal detection chain comprises a cascade of transistor-based low-noise amplifiers and mixer to translate THz signals to a lower, more manageable intermediate frequency (IF). In the THz regime, these components are increasingly noisier, less efficient, and sometimes even unavailable, leading to a substantial loss in signal-to-noise ratio (SNR) during the receiving process. The reported noise figures of state-of-the-art CMOS frontends deteriorate by roughly three orders of magnitude when scaling from 100 to 400 GHz, up to 40 dB[13,14]. This is mainly caused by intrinsic transistor degradation as the operating frequency approaches the transistor cut-off frequency (300-600 GHz, depending on transistor type)[14], as well as increasingly significant parasitic effects and degrading device quality factors. The latter two effects further limit the achievable antenna gain in chip-scale platforms, typically to below 5 dBi[15,16]. To push the transistor limit and mitigate the parasitic effects, sophisticated and band-specific architectures, as well as more advanced fabrication nodes are adopted, resulting in increased cost and complexity. Yet the systems see degraded and unsatisfactory noise performance at THz frequencies, as shown in **Fig. 1a**.

Leveraging broadband and low-loss photonic components, microwave photonic technology[17,18] holds the potential to overcome the bandwidth and loss limits of THz electronics and thus represents a compelling alternative to fulfill the THz promise[19,20]. Different from electronic frontends, antenna-captured signals in a photonic signal detection chain are firstly transduced into the optical domain using an electro-optic (EO) modulator, before further optical processing and/or low-loss distribution over optical fibers[21]. The processed signals are finally recovered at the target IF through either optical homodyne or heterodyne detection. The operation frequency range of this scheme, however, is typically limited to microwave bands due to the insufficient bandwidth of traditional bulk lithium niobate (LN) modulators. At even higher THz frequencies, photonic frontends usually necessitate additional discrete high-frequency electronic elements to first down-convert the THz signals to a modulator-accessible IF[22-26] before EO modulation, as shown in **Fig. 1b**. As a result, this architecture not only is bulky and costly, but more importantly is still plagued by the excessive noise in THz electronics. In recent years, the emergence of broadband integrated modulators beyond 100 GHz has opened opportunities for direct THz-optic conversion to unlock the full potential of THz photonics[27-30]. Initial attempts to further integrate EO modulators with on-chip antennas to realize full receiving frontends have also spanned multiple photonic platforms, including polymers[31-37], III/V materials[38,39], EO crystals[40] and thin-film lithium niobate (TFLN)[41,42]. However, most of these demonstrations operate below 100 GHz and typically suffer from limited antenna gain and unsatisfactory THz-optic conversion efficiency. Here, to accurately characterize and compare the free-space THz-optic conversion efficiencies across platforms, we define the THz-optic conversion figure of merit (FOM) as $FOM = CSR / P_{in}$, where $CSR$ is the carrier-to-sideband conversion ratio that captures the optical power conversion efficiency, which is then normalized by the illumination power density of free-space THz input, $P_{in}$. Based on this framework, we estimate that the FOMs of previously reported antenna-integrated THz-optic modulators have been limited to below 0.017 $cm^2/W$ (**Supplementary Table S1**). Given that any SNR loss in the detection chain cannot be recovered in subsequent amplification stages, the inefficient THz-optic conversion fundamentally precludes a full electrical-optical-electrical link with noise performance comparable to state-of-the-art electronics. To date, whether it is possible to break the fundamental THz link noise limitation, either through electronic or photonic approaches, has remained an outstanding question.

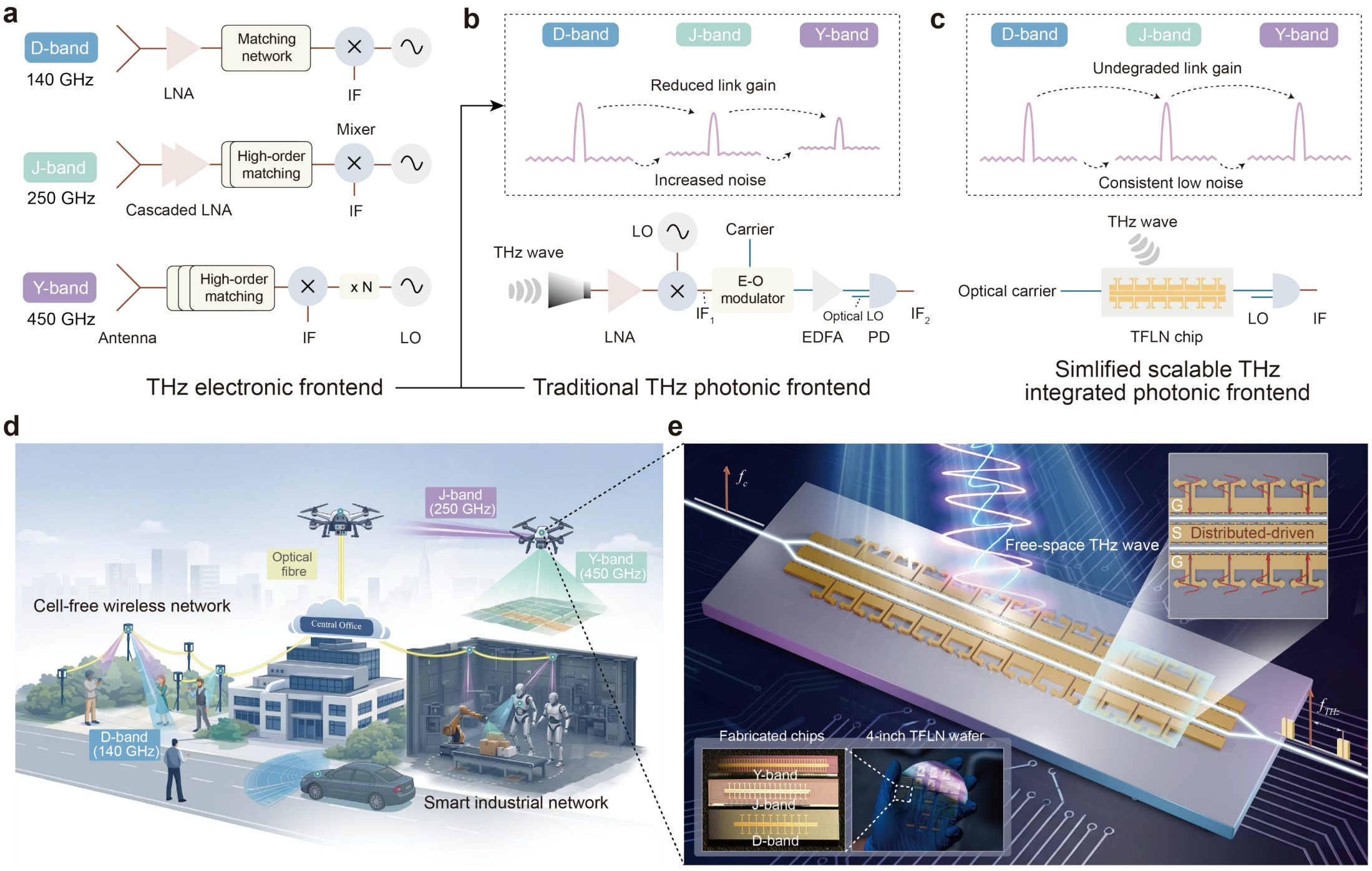

**Fig. 1 Integrated THz photonic frontend driven by antenna-integrated TFLN chip. a-c,** Schematic comparison of conventional THz electronic integrated frontends (a), a conventional THz photonic frontend (b), and the proposed THz photonic integrated frontend (c). Conventional THz electronic and photonic frontends see increasingly complicated architecture, increased noise, and reduced gain at higher frequencies, whereas the proposed solution offers a universal and simplified integrated architecture with undegraded performance at increased frequencies. The modulated optical output can be either directly processed on-site or relayed to a remote central office using low-loss fiber networks. LO, local oscillator; IF, intermediate frequency; LNA, low-noise amplifier; EDFA, erbium-doped fiber amplifier; PD, photodetector. **d,** Envisioned THz-ISAC scene enabled by the proposed integrated frontends, which provide a universal platform to leverage the band-specific THz advantages and enable compact, low-cost fiber fronthaul streaming to an edge cloud for cell-free and smart-industry wireless networks. **e,** Schematic of the proposed integrated frontends. An on-chip antenna array efficiently captures free-space THz signals and collectively drives a broadband EO modulator in a distributive manner, enabling continuous THz–optic interaction during signal reception. Inset shows photographs of the frontend chips with dimensions of 0.2 cm × 1.2 cm (D- and J-band) and 0.2 cm × 3.0 cm (Y-band), fabricated from 4-inch TFLN wafers.

In this article, we overcome these challenges by demonstrating integrated THz photonic frontends that, for the first time, deliver link noise performance superior to state-of-the-art electronics, empowered by a monolithic-integrated distributed-antenna-driven modulator architecture in the TFLN platform. Combining the ultrafast Pockels effect inherent to the LN crystal with enhanced EO interaction strength in the thin-film platform, TFLN modulators provide exceptional bandwidth, power consumption, and loss performances compared with other photonic platforms[30,43-45]. Our recent work characterized TFLN modulators at frequencies up to 500 GHz[30], where the modulation performance was mainly limited by the strong THz attenuation in metallic coplanar waveguides (CPW). Here, we adopt a new distributed receiving and modulation architecture, where basic broadband modulation elements and high-gain antenna units are co-designed and seamlessly co-integrated in a distributed manner. This enables a coherent and sustained free-space THz-to-optic conversion with minimal impact from on-chip THz attenuation. As a result, we achieve dramatically improved and frequency-scalable FOMs of 0.72, 0.54 and 0.52 $cm^2$/W across three major THz transmission windows[46], i.e., 140 GHz (D-band), 250 GHz (J-band), and 450 GHz (Y-band), respectively, representing improvements of over two orders of magnitude compared to previous records in the upper two windows. Combined with a shot-noise-limited heterodyne detection chain, the scalable THz-optic conversion over a 1.7-octave frequency span effectively translates to an undegraded link performance across the three bands. We verify this by measuring the effective isotropic noise figure (EINF), a system-level metric that jointly quantifies the added noise in both antenna reception and signal detection processes. Our results reveal improving EINF with frequency for a given optical insertion loss (IL), with measured EINFs as low as 13.6 and 16.2 dB in the J- and Y-bands, respectively, both outperforming state-of-the-art integrated electronics. The superior link EINF allows the complete elimination of high-frequency electronics and power-hungry optical amplifiers, leading to dramatically simplified and scalable architecture yet with uncompromised link noise performance, as shown in **Fig. 1c**. The performances of our integrated frontends in practical data transmission conditions are verified by a proof-of-concept 6G-oriented THz multi-link communication experiment, achieving aggregated data rates up to 20 Gbit/s with minimal hardware overhead.

## Results

**Figure 1d** shows an envisioned THz ISAC scene facilitated by our frequency-scalable integrated frontends. The low link noises across three major THz windows offer a universal platform to harness the unique advantages of specific THz bands, including D-band (~ 140 GHz, with low free-space loss) for high-precision radar, J-band (~ 250 GHz, with extensive available bandwidth) for high-speed communication, and Y-band (~ 450 GHz, with short wavelength) for high-resolution imaging. Moreover, the low cost, compact form factor and native fiber compatibility enable dense deployment and seamless fronthaul streaming to the edge cloud for centralized processing. These capabilities could potentially enable a paradigm shift toward cell-free wireless networks and wireless smart manufacturing, where dense distributed access points and centralized processing are essential[47-49]. **Figure 1e** depicts a schematic of the proposed THz integrated photonic frontend chip, which consists of an optical Mach–Zehnder interferometer (MZI), a CPW electrode and a high-gain dipole antenna array. The on-chip dipole antenna array efficiently captures free-space THz radiation and converts it into guided waves propagating along the CPW electrodes through periodic slots in the ground electrode. The antenna array is composed of a series of unit cells, each formed by two mirrored dipole antennas with respect to the central plane and excites the differential mode of the CPW electrode. An optical carrier is split into the two arms of the MZI by a multimode interference (MMI) power divider. Optical signals in the two arms then propagate through the slots formed by the signal and ground electrodes of the CPW, and are modulated by the differential THz fields in the slots, resulting in amplitude modulation when the two arms recombine. In this process, the modulator is collectively driven by the series of distributed dipole antenna elements, resulting in continuous and distributed interactions between the optical and THz waves as the latter is received. This distributed architecture is thus less prone to the substantial RF attenuation along the CPW line, especially at THz frequencies. The resulting modulation sidebands are subsequently heterodyned with a local oscillator (LO) in a photodetector (PD) to recover the signals at IF with shot-noise-limited sensitivity for further signal processing. Insets of **Fig. 1e** show photographs of the frontend chips in the three bands, fabricated through a 4-inch wafer-scale process (**Methods**).

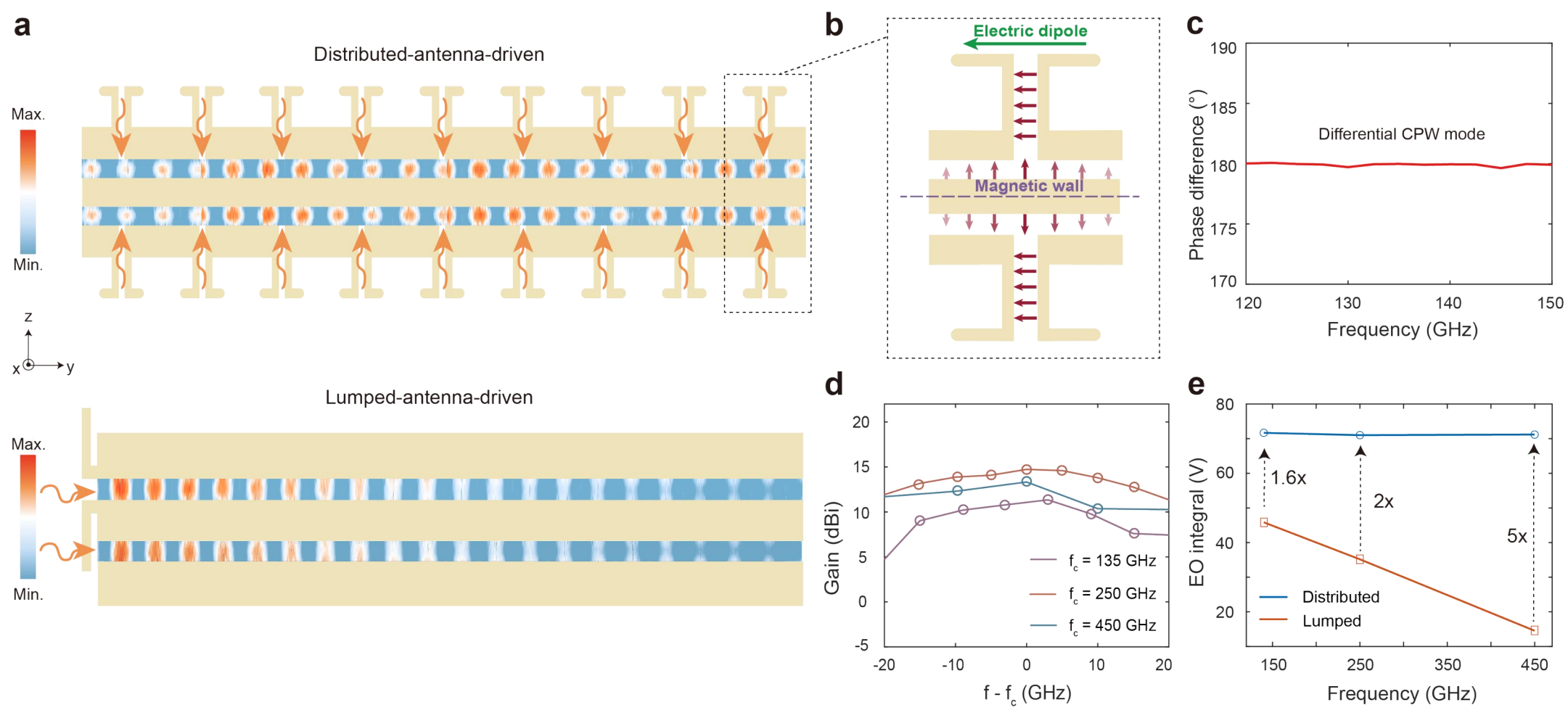

**Fig. 2 Design of distributed-driving antennas for frequency-scalable THz-optic interactions. a,** Schematic comparison of our distributed-antenna-driving architecture (top) and traditional lumped-antenna-driving architecture (bottom). The simulated distributions of the electric field magnitude inside the slots exhibit significant attenuation in the lumped-antenna case. The two simulations assume identical antenna gains and modulation lengths for fair comparison. **b,** Schematic of one THz antenna unit cell. Even CPW mode is excited with high purity by matching its symmetry with that of the antenna mode for push-pull modulation. Purple arrows illustrate the electric-field distributions. **c,** Simulated phase difference between the THz fields in the two slots under plane-wave excitations, indicating near-ideal differential mode excitation. **d,** Simulated antenna gains versus relative frequency in the three THz bands. **e,** Simulated EO integral versus frequency for distributed and lumped architectures.

To realize an efficient push-pull modulation with minimal phase skew, the differential CPW mode with even parity should be selectively excited, which features H-symmetry with respect to the waveguide central plane[50]. We achieve this using dipole antennas with a major polarization direction along the CPW propagation direction, i.e., the $y$-direction of LN crystal, as illustrated in **Fig. 2a-b**. Upon $y$-polarized plane-wave illumination, two in-phase electric dipoles aligned parallel to the central plane are excited, which also exhibits the H-symmetry and thus could efficiently excite the differential CPW mode with a strong suppression of the common mode, according to the even/odd mode selection rule. We validate this design rule by numerically simulating the phase differences between the THz fields in the two CPW slots. Taking our D-band chip based on TFLN-on-silicon substrate as an example, a nearly ideal $\pi$ phase difference is consistently achieved across a broad bandwidth from 120 to 150 GHz, as shown in **Fig. 2c**. Importantly, this is achieved without any additional transition structures. In contrast, lumped-antenna-driven architectures typically require baluns to excite the differential mode[41]. The baluns introduce additional loss and potential phase skew between the two MZI arms, which may result in compromised modulation efficiency and degraded signal fidelity[51].

To achieve high-gain signal reception, the on-chip antenna unit elements are further arranged into a one-dimensional array along the $y$-direction. With 10 elements and a period of 840 μm, our D-band on-chip antenna array achieves a highly directional radiation with a peak gain of 11.3 dBi at 138 GHz and a 3-dB gain-bandwidth exceeding 20 GHz, as shown in **Fig. 2d**. To fully exploit the broadband EO modulation capability of the TFLN platform, our J-band and Y-band designs leverage capacitively loaded CPW on quartz substrates, with which our previous results reveal effective modulation up to 500 GHz[30]. Benefiting from the advanced capacitively-loaded electrodes and low-permittivity quartz substrate, our J-band and Y-band on-chip antenna arrays achieve uncompromised gains of 14.3 and 13 dBi, respectively, ensuring undegraded system performance at increased frequencies (**Fig. 2d**). The detailed antenna dimensions and optimization process are provided in **Supplementary Notes 2-3**.

Importantly, the seamless connection between our dipole antenna and the traveling CPW mode allows the THz signals received by different elements of the array to immediately start interacting with the optical carrier in a distributed manner. This architecture ensures the THz-optic modulation process to constructively build up over a long device length with minimal impact from THz propagation loss, as the simulated THz field distributions in the CPW slots (top panel of **Fig. 2a**) indicate. In contrast, conventional lumped-antenna-driven designs typically feature separate antenna and modulation sections, leading to substantial THz wave attenuation in the transition and modulation sections (bottom panel of **Fig. 2a**). This ultimately limits the effective modulation length and the overall THz-optic conversion efficiency. For a fair comparison, the simulations in **Fig. 2a** assume identical antenna gain and CPW electrode parameters in the D-band**.** Further calculating the overall EO integrals[51] of the two schemes through full-wave simulations (**Methods**) reveals a 1.6× enhancement from our proposed distributed design in the D band, as shown in **Fig. 2e**. More importantly, the calculated EO integrals of our distributed-driven architecture show excellent frequency scalability, where the Y-band design features a minimal degradation of 0.1 dB from D band, despite a threefold increase in operating frequency (140 to 450 GHz). In contrast, lump-antenna-driven design exhibits a significant drop of EO integral with frequency due to increasingly significant THz loss (**Fig. 2e**). The undegraded EO interaction in our design effectively translates to up to 5× EO-integral improvement (or 25× higher power conversion efficiency) compared to lump-antenna-driven design in the Y-band, offering a practical solution to overcome the performance degradation at high frequencies in both electronics and traditional photonics schemes.

We experimentally characterize the free-space THz-optic response of our device using the experimental setup shown in **Fig. 3a**. The THz signals are generated using frequency multipliers that upconvert the input signal from a microwave source to 115-150, 220-280 and 430-460 GHz for our three chips, respectively. The multiplied signals are then delivered to the devices through far-field free-space emission using standard-gain horn antennas. **Figure 3b** shows the measured optical spectra of the modulated sidebands with spacing from the carrier in agreement with the THz source frequencies. This confirms the direct free-space THz-optic conversion by our THz photonic chips. **Figure 3c** illustrates the measured CSRs at various incident angles and THz frequencies. The CSRs are normalized to a fixed free-space power density of 10 $W/m^2$ to account for frequency-dependent variations in the THz source output, such that the panels in **Fig. 3c** could capture and directly compare the chips' intrinsic frequency and angular responses. Results from all three chips show consistent angle-frequency correlations that match well with the beam-scanning characteristics of on-chip antenna arrays as well as theoretical predictions (dashed white curves).

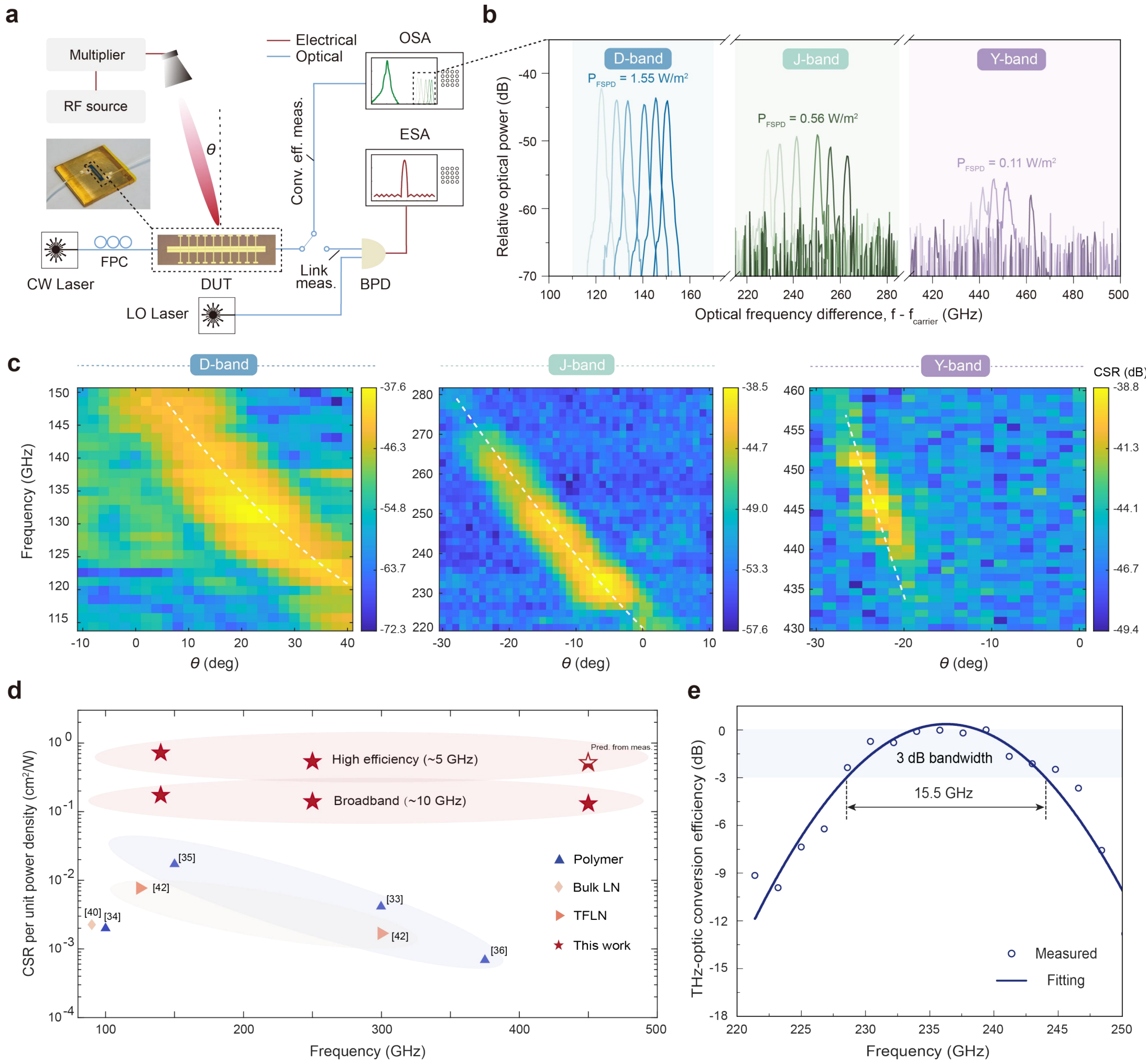

**Fig. 3 Free-space THz to optic conversion. a,** Schematic of the experimental setup for CSR measurement, where free-space THz signals are generated by a THz multiplier and emitted to the chip by a horn antenna. The resulting modulated optical signals are either captured by an optical spectrum analyzer for CSR measurement or heterodyned with a LO laser in a balanced photodetector for link performance analysis in an electronic spectrum analyzer. FPC, fiber polarization controller; CW, continuous wave; DUT, device under test; OSA, optical spectrum analyzer; BPD, balanced photodetector; ESA, electronic spectrum analyzer. **b,** Measured optical spectra of modulated sidebands across the D-, J- and Y-bands. FSPD, free-space power density. **c,** Angle-frequency map of the measured CSRs in the three bands. To capture the chip's intrinsic frequency and angular responses, the CSRs are normalized to an input FSPD of 10 W/m$^2$ to de-embed the power variations of the THz sources in different bands. Dashed curves correspond to the theoretically predicted angle-frequency relation. **d,** Comparison of the figures of merit with previous demonstrations. **e,** Relative THz-optic conversion efficiency as a function of frequency for a fixed incident angle of -9° for the J-band chip, showing a 15.5 GHz 3-dB bandwidth.

Compared to previous demonstrations, our integrated frontends effectively extend the performance boundary in terms of both conversion efficiency and frequency scalability, as shown in **Fig. 3d**. Our chips achieve CSRs of -37.6, -38.5 and -38.8 dB under a normalized free-space power density of 10 W/m$^2$, at 134, 250 and 450 GHz, respectively, which translates to FOMs of 0.17, 0.14 and 0.13 cm$^2$/W (bottom red stars in **Fig. 3d**). These values not only represent at least one order of magnitude improvement over all the previous results (**Fig. 3d**; detailed comparison can be found in **Supplementary Table S1**), but more importantly shows nearly undegraded performance across the three THz bands, i.e. 1.2 dB degradation over 1.7 frequency octaves. The significant improvement stems from a unique combination of the high-gain THz antenna design, the distributed driving architecture, and strong and broadband EO interaction in the TFLN platform. In contrast, the THz-optic conversion efficiencies in previous demonstrations typically see significant degradation with frequency, due to limited antenna gains and their lump-antenna-driven nature. Notably, the high and scalable conversion efficiencies are achieved with consistent available bandwidth. Due to the angle-frequency correlation (**Fig. 3c**), the THz-optic efficiency drops as it moves away from the optimal frequency, as shown in **Fig. 3e**. Nevertheless, our D-band, J-band and Y-band chips all offer broad and consistent bandwidths around 10 GHz (**Supplementary Fig. S3**), sufficient for high-speed THz communications with tens of Gbit/s data rate. This bandwidth could be further extended to 15.5 GHz (**Fig. 3e** and **Supplementary Note 4**) at a different incident angle while maintaining a high FOM of 0.07 $cm^2/W$. In addition, the high conversion efficiency and broad bandwidths are consistently achieved within angular ranges from 13° to 29° for the D-band chip and -16° to -5° for the J-band chip (**Supplementary Note 4**). The consistent performance across a broad angular range is crucial for multi-link high-speed THz communications, as will be discussed later. Applying devices with even longer interaction lengths of 26 mm, we can further improve the FOMs by nearly 4 times to 0.72 and 0.54 $cm^2/W$ for our D-band and J-band designs (**Supplementary Note 5**), respectively, while maintaining a bandwidth around 5 GHz (top red stars in **Fig. 3d**). Applying the same measured scaling law to the Y-band chip design leads to a predicted FOM of 0.52 $cm^2/W$ (half-filled red star in **Fig. 3d**). These results again show the excellent scalability across bands, with more than two orders of magnitude improvement over previous records in the J- and Y-bands.

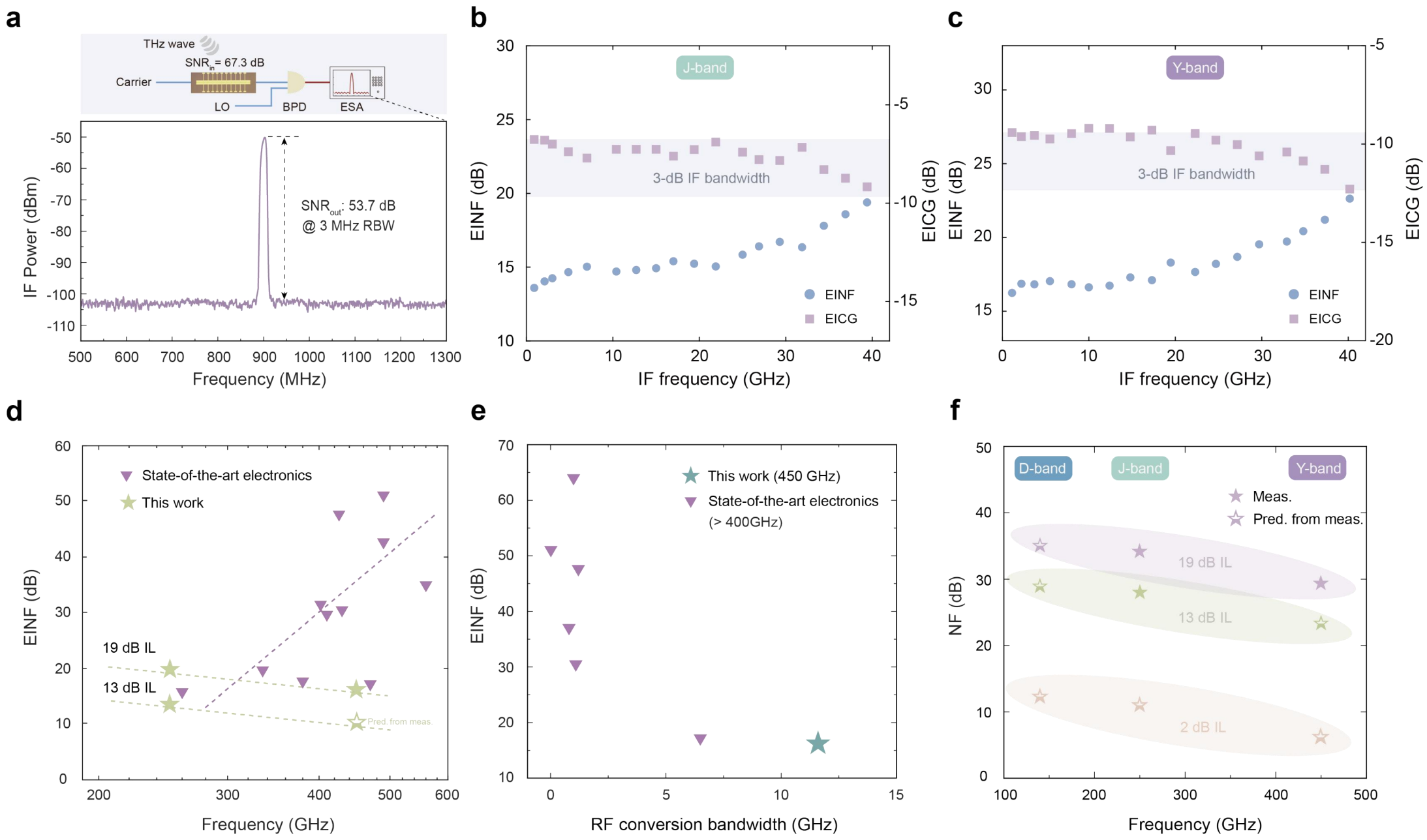

**Fig. 4 Low-noise photonic THz receiving link outperforming electronics. a,** Measured output IF spectrum showing a SNR of 53.7 dB for a 250 GHz free-space input signal with an input SNR of 67.3 dB, yielding an EINF of 13.6 dB. **b-c,** Measured EINF (blue) and EICG (purple) versus IF, at 250 GHz (b) and 450 GHz (c), both showing a broad 3-dB IF bandwidth around 40 GHz. **d,** Measured (solid stars) and calculated (half-solid star and green dashed lines) ENIFs in comparison to state-of-the-art electronics (purple), revealing a decreasing EINF with frequency for a given optical IL. **e,** Performance comparison of EINF and RF conversion gain bandwidth (green) with state-of-the-art electronics (purple) operating above 400 GHz. **f,** Measured (solid stars) and predicted (half-solid stars) NFs in the three THz bands under various ILs.

Based on the scalable high-efficiency THz-optic conversion process, we demonstrate a low-noise THz frontend link that outperforms state-of-the-art electronics yet with a greatly simplified architecture. In this experiment, the THz incident angles are tuned to match the optimal values determined by the measured CSR patterns in **Fig. 3c**. After THz-optic conversion, the generated optical sidebands from the chips are heterodyned with a LO laser in a balanced photodetector (BPD) to recover the signals at IFs from 1-40 GHz (**Methods**). With an input free-space SNR of 67.3 dB in the J-band (**Methods**), our full detection link produces an output SNR of 53.7 dB (**Fig. 4a**), yielding an EINF (i.e., SNR degradation through the entire link) as low as 13.6 dB (**Methods**) around 1 GHz IF frequency. **Figure 4b-c** shows the measured effective isotropic conversion gain (EICG) and EINFs at different IFs for J- and Y-bands, respectively. Results indicate consistent 3-dB bandwidths around 40 GHz for both bands, which are mainly limited by the 3-dB bandwidth of the BPD. The measured EINF in the Y-band (at 1 GHz IF) is 16.2 dB, which slightly increased from that in the J-band due to a 6 dB higher optical IL, likely due to one-time fabrication imperfection. Nevertheless, both measured EINF values here have outperform state-of-the-art integrated THz frontends in their respective bands (**Fig. 4d**, see detailed comparison in **Supplementary Table S3**). In contrast to electronic frontends whose EINF inevitably degrades with frequency, our integrated photonic frontends exhibit positive performance scaling with frequency, as revealed by the decreased simulated and measured EINF trends (green dashed lines in **Fig. 4d**) assuming constant ILs. This favorable performance scaling stems from the scalable FOM described above (**Supplementary Note 6**) and is in stark contrast to the typical frequency scaling law in electronic frontends (purple dashed line in **Fig. 4d**). **Figure 4e** further summarizes the RF conversion gain bandwidth and EINF performances of state-of-the-art integrated frontends above 400 GHz. With an 11.6 GHz 3-dB conversion bandwidth (**Supplementary Note 7**) and a superior EINF, our Y-band frontend effectively outperforms state-of-the-art electronics in both key metrics. By de-embedding the simulated antenna gain $G$ from the measured EINF, we further evaluate the equivalent link noise figure (NF) of our frontends according to: $NF = EINF + G$, which amounts to 27.9 dB and 29.2 dB at 250 and 450 GHz, respectively (solid stars in **Fig. 4f**). Currently, the achieved NF and EINF values are mainly limited by the relatively large optical ILs of our chips, including both external fiber-chip-coupling (5 ~ 7 dB/facet) and on-chip losses. Assuming a substantially lowered IL of 2 dB, which is achievable by employing an optimized dry etching process[52], double-layer edge couplers[53], and/or heterogeneously integrated PDs[54,55], we predict ultralow NFs of 12.1 dB, 11.1 dB, and 6.2 dB at 140, 250 and 450 GHz (**Fig. 4f** and **Supplementary Table S4**). These NF values of the detection links themselves would be on par with state-of-the-art electronics in the D-band and J-band, and outperform those in the Y-band (**Supplementary Fig. S8**). Further combining them with the high-gain antennas developed in this work could lead to negative EINFs for the full receiving frontend (**Supplementary Table S3**), a regime inaccessible to current integrated electronic solutions. Overall, our integrated photonic frontends not only provide link noise performance that, for the first time, outperforms electronics frontends, but also feature a greatly simplified architecture and excellent frequency scalability.

Finally, we verify the real-world application capability of our THz photonic frontend by demonstrating a high-speed THz photonic wireless communication link. Free-space THz data stream from an electronic transmitter 200 mm away is collected by our integrated photonic frontend and down-converted to IF band by heterodyning the generated optical sidebands with a LO laser 10 GHz away from the sideband (**Fig. 5a**). We successfully achieve high-speed wireless data transmission up to 10 Gbit/s, with measured bit-error ratios (BERs) below $2.5\times10^{-3}$ (the inset on the right side of **Fig. 5a**), showing accurate baseband-signal retrieval without relying on any high-frequency electronic components.

Beyond single-lane wireless communications, we further show our integrated frontend can serve as a key enabler for multi-link communications to fulfill the cell-free vision in 6G networks, in which the ability to interface multiple concurrent links is required for blockage resilience and multi-base-station collaboration in the THz band[56,57]. As a proof of concept, we first demonstrate our integrated frontend can support two concurrent, high-speed and independent downlinks from two base stations, emulated by two electronic transmitters with different allocated carrier frequencies (129.6 and 135.6 GHz) and angular positions (24° and 18°) (**Fig. 5b**), following the angle-frequency characteristics of our integrated frontend (**Fig. 3c**). To showcase the independent operation of the two links, the two carriers were respectively modulated by 4-ary quadrature amplitude modulation (4QAM) and quadrature phase shift keying (QPSK) formats. Our integrated frontend allows the simultaneous down-conversion of the received THz signals by heterodyne detection with an optical LO that is 118.5 GHz away from the optical carrier, generating two IF signals centered at 11.1 and 17.1 GHz, respectively, as shown in **Fig. 5c**. The undesired spikes at the two center frequencies arise from LO leakage in the electronic transmitters and can be mitigated using photonics-based THz carrier generation[58]. The clearly defined constellation diagrams of the retrieved data in **Fig. 5d** reveal consistent, reliable and simultaneous data transmission for the two links, both achieving up to 10 Gbit/s wireless communication with BER below the soft-decision forward-error correction (SD-FEC) threshold. Further comparison of the constellation diagrams with the single-link case (either link 1 or 2 is turned on) in **Fig. 5d-e** indicates multilink operation results in only a marginal increase in BER. Notably, this is achieved without additional hardware consumption compared to the single-link case.

Leveraging coherent multi-link reception, we further show the application potential of our integrated frontend for boosting channel capacity and enhancing channel quality to meet the demands of high-throughput 6G network. Specifically, we perform multi-link carrier aggregation using two concurrent QPSK-modulated data streams, effectively achieving a doubled data rate up to 20 Gbit/s with a BER lying between the two individual channels and remaining below the SD-FEC threshold, as shown in **Fig. 5f.** Given the wide operation bandwidth of our integrated frontend, we anticipate that a single frontend chip can simultaneously interface at least 4 high-speed links, with aggregate data rates of 40 Gbit/s. Utilizing a similar configuration, modulating both carriers with identical baseband signals and applying maximum ratio combining (MRC) in the digital signal processing (DSP) yields a substantial BER reduction from $7.7\times10^{-3}$ to $2.9\times10^{-3}$ (**Fig. 5g**). Similar aggregate data rates and multi-link operation are also readily achievable in the J-band and I-band based on the respective frontend implementations discussed above. With the demonstrated versatile multi-link interconnectivity and low link noises, our integrated photonic frontends hold significant potential for scalable high-throughput THz multi-access communication systems in the 6G era.

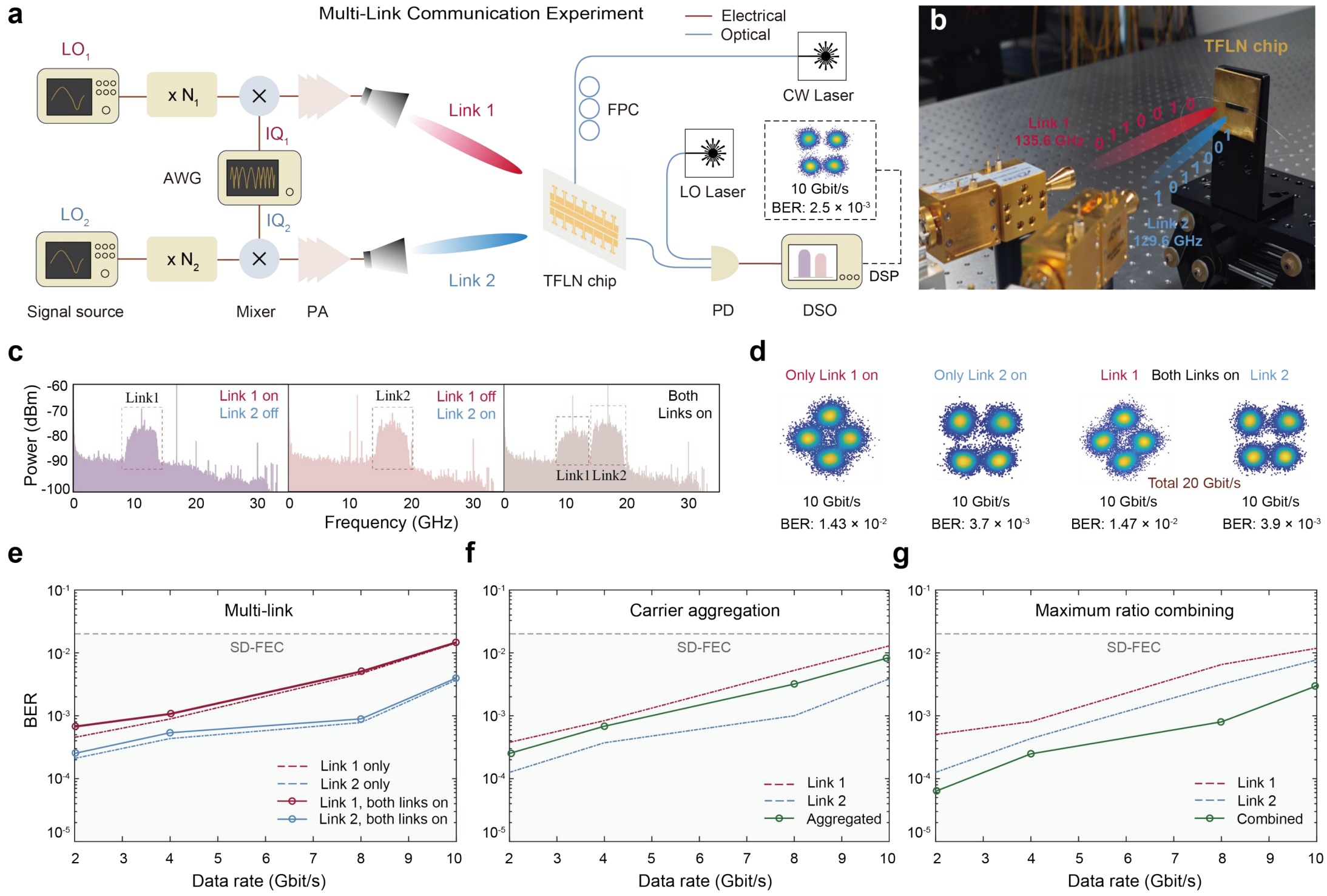

**Fig. 5 Multi-link high-speed THz communications. a,** Schematic illustration of the experimental setup, where the integrated frontend simultaneously interfaces two coexisting links with different angular positions (24° and 18°). The received THz signals are down-converted to IF in parallel and demodulated concurrently. PA, power amplifier; AWG, arbitrary waveform generator; DSO, digital storage oscilloscope; DSP, digital signal processing. **b**, Photo of the experimental setup, where two identical electronic transmitters and horn antennas are utilized to establish two coexisting links. **c,** IF spectra for the single-link (left and middle) and multi-link (right) cases. **d,** Constellation diagrams of the retrieved baseband data for the multi- and single-link cases. **e**, Bit-error ratios (BER) for the single-link (dashed) and multi-link (solid) cases. **f.** BERs before (dashed) and after (solid) carrier aggregation. **g,** BERs before (dashed) and after (solid) maximum ratio combining (MRC). Grey dashed lines in (e-g) indicate soft-decision forward-error correction (SD-FEC) threshold.

## Discussion

We have demonstrated integrated THz photonic frontends that deliver consistently low-noise receiving links across three major THz windows in a universal and simplified architecture. These unique features are enabled by the seamless co-integration of broadband TFLN EO modulator and high-gain on-chip THz antenna array in a distributive-driven architecture. We achieve frequency-agnostic free-space-THz-to-optic conversion with FOMs orders of magnitude higher than previous records. As a result of this enhanced THz-optic conversion, we attain link EINFs of 13.6 and 16.2 dB at 250 and 450 GHz, respectively, both surpassing state-of-the-art integrated electronics in their respective bands.

Our demonstrated link noises, currently limited by the substantial optical IL, could be further improved by at least 10 dB based on state-of-the-art TFLN fabrication techniques and double-layer fiber-to-chip edge couplers[53]. Assuming an IL of 2 dB and a higher optical pump power of 30 dBm[59,60], we anticipate an unprecedented link NF below 5 dB (EINF of -9.8 dB) at 450 GHz (**Supplementary Tables S3 and S4**), significantly surpassing the current state of the art. Beyond loss engineering, additional system-level headroom exists to increase the lane speed. Specifically, the baud rate could be significantly improved by operating in the J-band to increase the usable bandwidth (**Fig. 3e**). Incorporating high-gain THz polytetrafluoroethylene lenses (30 dBi)[61] at both the transmitter and receiver sides could compensate for the significant THz free-space loss in long-distance links. These optimizations could collectively lead to THz wireless links with ultra-high speed (>100 Gbit/s) and long reach (>100 m) without compromising the cost and weight advantage of our integrated platform.

With the greatly simplified architecture, our photonic frontends pave the avenue for fully integrated THz photonic receivers. Specifically, the BPD involved in our frontends can be heterogeneously integrated on our TFLN chip [62].The core photonic chip could be further co-packaged with the LO and carrier laser chips through photonic wire bonds [63], which will not only significantly reduce system size, weight and power consumption (SWaP), but also further extend the performance boundary by eliminating all the interconnection losses in our current setup.

Finally, leveraging the low optical loss, excellent scalability, and wafer-scale manufacturability of the TFLN platform, our integrated THz photonic frontend architecture could be further scaled up to 2-D arrays with even higher THz-optic conversion efficiency and dynamic beam steering capability. By integrating on-chip phase shifters between antenna elements and/or tunable delay lines, such chip-scale phased-array frontends with agile beam-scanning ability and dynamic environmental adaptability could become critical enabler of ultra-compact, adaptive and scalable 6G ISAC systems.

## Methods

### Device fabrication

Devices are fabricated on a commercial *x*-cut LN wafer (NANOLN) on a silicon substrate for the D-band devices and on a quartz substrate for the J- and Y-band devices with a 500-nm-thick LN thin film and a 2-μm buried $SiO_2$ layer. A $SiO_2$ layer is first deposited on the surface of the 4-inch LNOI wafer via plasma-enhanced chemical vapor deposition (PECVD) to serve as an etching hard mask. Waveguides are patterned on the entire wafer using an ASML UV Stepper lithography system (NFF, HKUST). The die size is 1.5 cm × 1.5 cm, with a resolution of 500 nm. The exposed resist patterns are first transferred to the $SiO_2$ layer via a standard fluorine-based dry etching process. The waveguide patterns are then transferred into the TFLN by dry etching 250 nm of the LN film using an optimized Argon-plasma ($Ar^+$) based inductively coupled plasma reactive ion etching (ICP-RIE) process. The fabricated rib waveguides have a height of 250 nm and a top width of 1.2 μm, respectively. After removing the residual $SiO_2$ mask and redeposited materials, an annealing process is performed to improve structural stability. A second stepper lithography, metal evaporation, and lift-off process are employed to fabricate the electrodes. The chip edges are finally polished to enhance fiber-chip coupling.

### Numerical simulation

The numerical simulations of THz devices are performed using high frequency structure simulator (HFSS) based on finite element method (FEM). In the simulation, *y*-polarized plane waves are applied to excite the devices, mimicking free-space waves in the actual scenarios. The phase change induced by the EO modulation depends linearly on the local modulating electric field intensity $E(x,t)$ and can be calculated as[51]:

$$\Delta\varphi = \frac{\pi}{\lambda} n^3 r_{33} \Gamma \int_0^L KE(x, \frac{x}{V_o}) dx$$

where $r_{33}$ is the EO coefficient, $\Gamma$ is the overlap integral, $n$ is the optical refractive index, $K$ is a conversion constant between voltage and electric field intensity, and $V_o$ is the optical group velocity. Note that, except for $E(x,t)$, all the other conditions are identical for both the lump-antenna-driven and proposed distributed-antenna-driven cases. Therefore, we calculate the modulation field integral under the same free-space power density to quantitatively characterize the impacts of the field profiles on THz-optic conversion efficiency and verify the architectural advantage. As the distributive-driven architecture is less prone to substantial THz loss, it promises the direct scaling of the THz-optic interaction strength with the effective interaction length. Meanwhile, our wafer-scale manufacturability on the TFLN platform provides an ideal toolbox for fabricating devices with different feature sizes. Combining both the architectural advantage and the fabrication flexibility, we extend the effective interaction length to 28 mm in the Y-band design to deliver almost consistent THz-optic interaction strength across the three bands, as verified by the minimal EO integral degradation from 140 to 450 GHz in Fig. 2e.

**Characterization of the free-space THz-optic response**

The THz signal is generated using frequency multipliers (Millitech AMC-08-RFH00 and VDI SGX series) that upconvert the input signals from an RF source to 115-150, 220-280 and 430-460 GHz for our D-, J- and Y-band measurements, respectively. For D-band chip characterization, we used both Millitech AMC-08-RFH00 and VDI WR5.1 SGX to cover the 115-150 GHz range. To accurately retrieve the intrinsic THz-optic response, the multiplier output power was carefully measured using an RF power meter (Erickson PM5B) prior to measurement. The multiplied signals are then delivered to standard-gain horn antennas, which emits free-space THz waves onto the device. The chip is positioned 150 mm away from the horn antenna, providing an emulation of the plane wavefront in a real wireless communication scenario. The THz–optic response was characterized in the telecom C-band using a tunable continuous-wave laser (Santec TSL-550) as the optical carrier. Before entering the chip, a three-paddle fiber polarization controller (FPC) aligns the input light's polarization to the transverse-electric (TE) mode, ensuring the optimal EO interaction strength in LN. The on-chip Mach–Zehnder modulator (MZM) is biased at the quadrature point to accurately extract the carrier-to-sideband conversion ratio. The chip output optical spectrum is recorded by an optical spectrum analyzer (Yokogawa AQ6370).

**Characterization of THz frontend link noise**

The THz frontend link was characterized using the same free-space THz generation and illumination configuration described in the previous section. The chip is pumped by a narrow-linewidth continuous-wave laser (ConLAS CaSF-D) with an on-chip pump power of 125 mW. The polarization states of the lasers were adjusted using three-paddle fiber polarization controllers to excite the waveguide's fundamental TE mode. Optical coupling to and from the waveguide facets of the TFLN chip was accomplished using tapered lensed fibers. The MZM is biased near its null point to improve the conversion gain[64] while minimizing the carrier noise. A second telecom tunable laser (Santec TSL-550) served as the local oscillator (LO) for optical heterodyne detection. The chip output and the LO laser were then combined with an optical hybrid and fed into a balanced photodetector (OVLINK BPD-40-SM-FA) for optoelectronic down-conversion. A LO power of 6 dBm was used to ensure the link is shot-noise-limited. The on-chip antenna array converts the free-space THz signals into on-chip THz signals, which subsequently modulate the optical carrier. The chip's optical output was combined with the LO laser in a 50/50 fiber coupler and fed to the BPD. The resulting intermediate frequency (IF) beat note between the LO and the THz-induced sideband was captured by a broadband electrical spectrum analyzer (Rohde & Schwarz FSV3044) with a 44 GHz analysis bandwidth. The insertion losses of the RF coaxial cables are measured and de-embed to accurately capture the device's intrinsic response. The effective isotropic conversion gain (EICG) was calculated as:

$$EICG\ (dB) = P_{IF} - P_r$$

where $P_{IF}$ and $P_r$ are the measured IF output power and the received power of an isotropic receiving antenna, respectively. The latter is given by Friis transmission equation:

$$P_r = P_t + G_t - 20\log(\frac{4\pi df}{c})$$

where $P_t$ is the multiplier output power, $G_t$ is the gain of the transmitting horn, $d$ is the free-space propagation distance, $f$ is the THz frequency and $c$ is the speed of light. The difference between $P_r$ and thermal noise floor gives the input signal-to-noise ratio (SNR). For example, in Fig. 4a, with 0 dBm multiplier output power, 22 dBi transmitting horn gain, 15 cm distance and 3 MHz bandwidth, the input SNR is 67.3 dB.

The ratio between the input SNR and the measured output SNR then determined the effective isotropic noise figure (EINF):

$$EINF(dB) = P_N(dBm/Hz) - 174(dBm/Hz) - EICG$$

where $P_N$ is the measured output noise floor. Thermal noise of -174 dBm/Hz is assumed for the input signal.

The equivalent noise figure (NF) of the link (an antenna-less system), considering a separate antenna and signal detection chain was obtained from:

$$NF(dB) = EINF + G_r$$

where $G_r$ is the gain of the on-chip antenna array.

**THz communication experiment.**

To implement the THz wireless transmission experiments, two sets of electronic-based transmitters are adopted. Specifically, two independent radio frequency tones at 21.6 GHz and 22.6 GHz are generated using signal generators. Each tone is fed into a frequency multiplier (6×) to produce THz signals at 129.6 GHz and 135.6 GHz, respectively. The two THz signals are then supplied to separate IQ mixers. A 64 GSa/s arbitrary waveform generator (AWG, Keysight M8195A) provides baseband signals in 4-ary quadrature amplitude modulation (4QAM) or quadrature phase shift keying (QPSK) format at symbol rates of 1, 2, 4, or 5 GBaud to drive the I and Q ports of each mixer. The baseband QPSK/4QAM signals are pulse-shaped with a root-raised cosine filter with a roll-off factor of 0.3. Baseband timing alignment between the four AWG outputs was calibrated prior to measurements to minimize co-channel interference. The upconverted outputs from the IQ mixers are amplified by power amplifiers (PAs) before being emitted toward the chip. The on-chip MZM is biased near its null point to generate a carrier-suppressed double-sideband signal. The LO laser is positioned near the upper sideband of the chip output signal. The LO laser and chip output signals are combined and detected by a photodetector (Conquer KG-BPR 40G). The IF signal after optical detection is recorded by an 80 GSa/s real-time oscilloscope (Keysight DSOX93204A) and processed offline using digital signal processing (DSP) algorithms.

## Funding

This work is supported by the National Natural Science Foundation of China (62525507, C.W., 62471433, X.B.Y., and 625B2157, Z.X.W.), Research Grants Council, University Grants Committee (STG3 E-104-25-N, C.W., CityU 11204022, C.W., CityU 11213125, C.W., and JRFS2526-1S01, H.F.), Croucher Foundation (9509005, C.W.), City University of Hong Kong (9610682, C.W.), Department of Science and Technology of Guangdong Province (Guangdong and Hong Kong Universities '1+1+1' Joint Research Collaboration Scheme, H.F.), Natural Science Foundation of Zhejiang Province (9509005, X.B.Y.).

## Author contributions

Y.S.Z., Z.X.W. and C.W. conceived the idea. Y.S.Z. and Z.X.W designed the devices and performed numerical simulations with the discussions from S.W.Q. and J.H. The chips were fabricated by Y.F.W., H.F. and Z.X.C.; Y.S.Z, Z.X.W. and Y.W.Z. carried out the THz-optic characterization measurements. Y.S.Z., Z.X.W. and Y.S.T. conducted link measurements with the assistant from X.Z.X. The THz communication experiments were performed by L.B. and Z.D. with the guidance from L.Z. and X.B.Y. The manuscript was prepared by Y.S.Z. and Z.X.W. with contributions from all the authors. C.W., X.B.Y. and C.H.C. supervised the project.

**Acknowledgments.**

We thank Dr. W.-H. Wong, Dr. K. M. Shum and Dr. K. F. Chan at CityU for the help in device fabrication and measurements. We thank the technical support of C. F. Yeung, S. Y. Lao, C. W. Lai and L. Ho at HKUST, Nanosystem Fabrication Facility (NFF), for the stepper lithography and PECVD process.

**Competing interests**

H.F., Z.X.C. and C.W. are involved in developing lithium niobate technologies at RhinoptiX Technology Ltd.

**Data availability.**

Data underlying the results presented in this paper are not publicly available at this time but may be obtained from the authors upon reasonable request.

## Supplementary Information

### Supplementary Note 1: Details of the comparison with the state-of-the-art photonic frontends

Current antenna-integrated THz photonic frontends exhibit limited THz-optic conversion with FOM below 0.017 $cm^2/W$, as detailed in **Supplementary Table S1**. Moreover, they show significant degradation with operating frequency. Our integrated frontends effectively extend the performance boundary in terms of both conversion efficiency and frequency scalability.

**Supplementary Table S1 Performance comparisons of the state-of-the-art antenna-integrated modulators**

| Ref. | Platform | Operating frequency (GHz) | Free-space power density ($W \cdot m^{-2}$) | CSR (dB) | FOM ($cm^2/W$) |
|---|---|---|---|---|---|
| [40] | Bulk LN | 90 | 443 | -40 | 0.0022 |
| [34] | Polymer | 100 | 12.8 | -56 | 0.0019 |
| [35] | Polymer | 150 | 34.3 | -42.3 | 0.017 |
| [33] | Polymer | 300 | 15.7$^{\dagger}$ | -51.8 | 0.0041 |
| [36] | Polymer | 375 | 290 | -47 | 0.00068 |
| [42] | TFLN | 125 | 12995.5$^{\ddagger}$ | -20 | 0.0076 |
| | | 300 | 1875.8$^{\ddagger}$ | -35 | 0.0016 |
| **This work** | TFLN (broadband chips) | 133.6 | 1.55 | -45.7 | **0.17** |
| | | 250 | 0.56 | -51.1 | **0.14** |
| | | 450 | 0.11 | -58.3 | **0.13** |
| | TFLN (high-efficiency chips) | 126.4 | 1.98 | -38.5 | **0.72** |
| | | 247 | 0.34 | -47.3 | **0.54** |
| | | 450 | - | - | **0.52$^{\S}$** |

$^{\dagger}$ Estimated assuming a typical lens gain of 20 dB

$^{\ddagger}$ Calculated based on the measured on-chip power coupling efficiency

$^{\S}$ Predicted from measured results

### Supplementary Note 2: Detailed parameters of the on-chip antenna elements

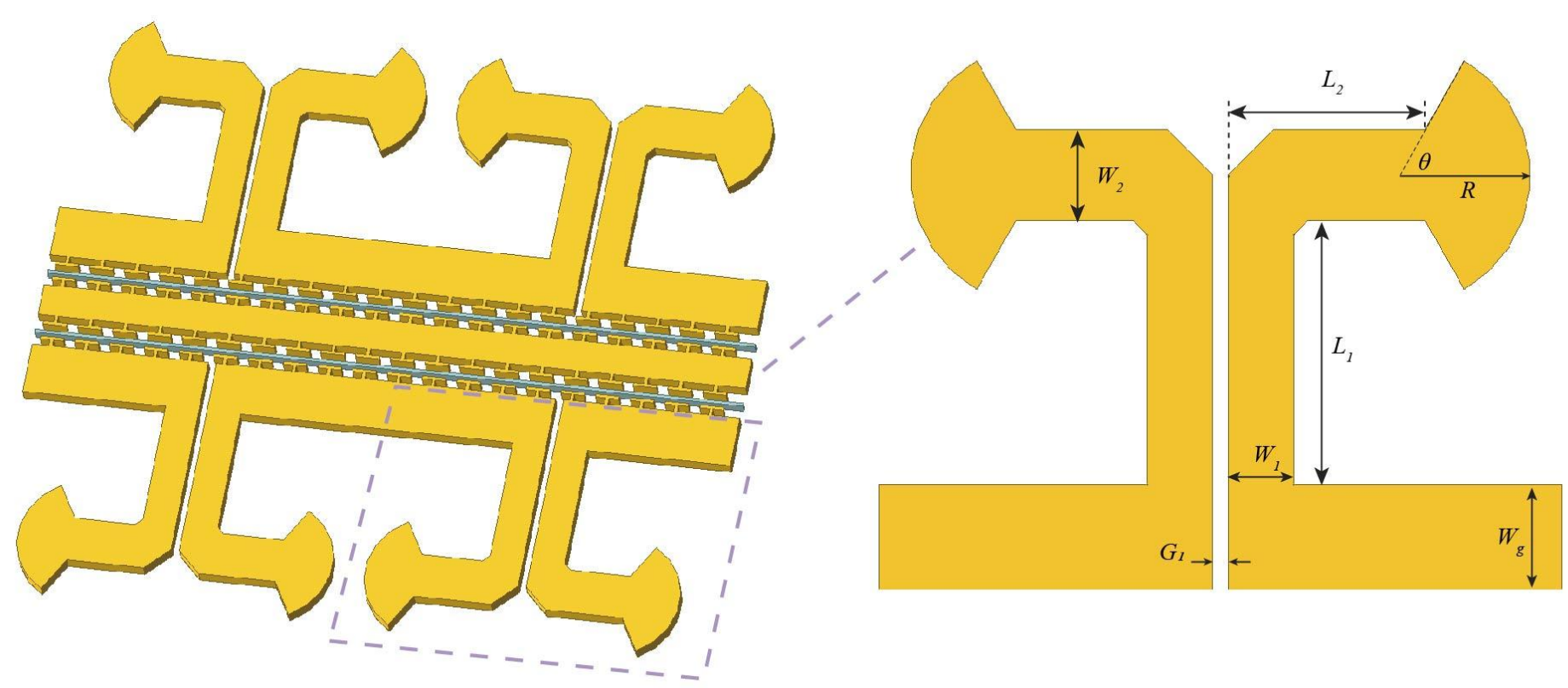

**Supplementary Fig. S1 | Detailed antenna dimensions**

The detailed dimensions of the on-chip antennas are summarized in **Supplementary Fig. S1** and **Supplementary Table S2**. To fully exploit the TFLN modulator's broad bandwidth, the J- and Y-band designs use a capacitively loaded CPW on a quartz substrate. Each dipole arm employs a sector-shaped end-loading structure to expand the operation bandwidth.

**Supplementary Table. S2 Detailed dimensions of the on-chip antenna elements (unit: µm)**

| D-band | | | | | | | |
|---|---|---|---|---|---|---|---|
| $W_g$ | $W_1$ | $W_2$ | $L_1$ | $L_2$ | $G_1$ | $\theta$ | $R$ |
| 150 | 50 | 100 | 650 | 250 | 6 | 90° | 50 |

| J-band | | | | | | | |
|---|---|---|---|---|---|---|---|
| $W_g$ | $W_1$ | $W_2$ | $L_1$ | $L_2$ | $G_1$ | $\theta$ | $R$ |
| 150 | 50 | 70 | 460 | 100 | 5.5 | 60° | 80 |

| Y-band | | | | | | | |
|---|---|---|---|---|---|---|---|
| $W_g$ | $W_1$ | $W_2$ | $L_1$ | $L_2$ | $G_1$ | $\theta$ | $R$ |
| 150 | 50 | 70 | 300 | 110 | 5.5 | 60° | 80 |

## Supplementary Note 3: On-chip antenna optimization

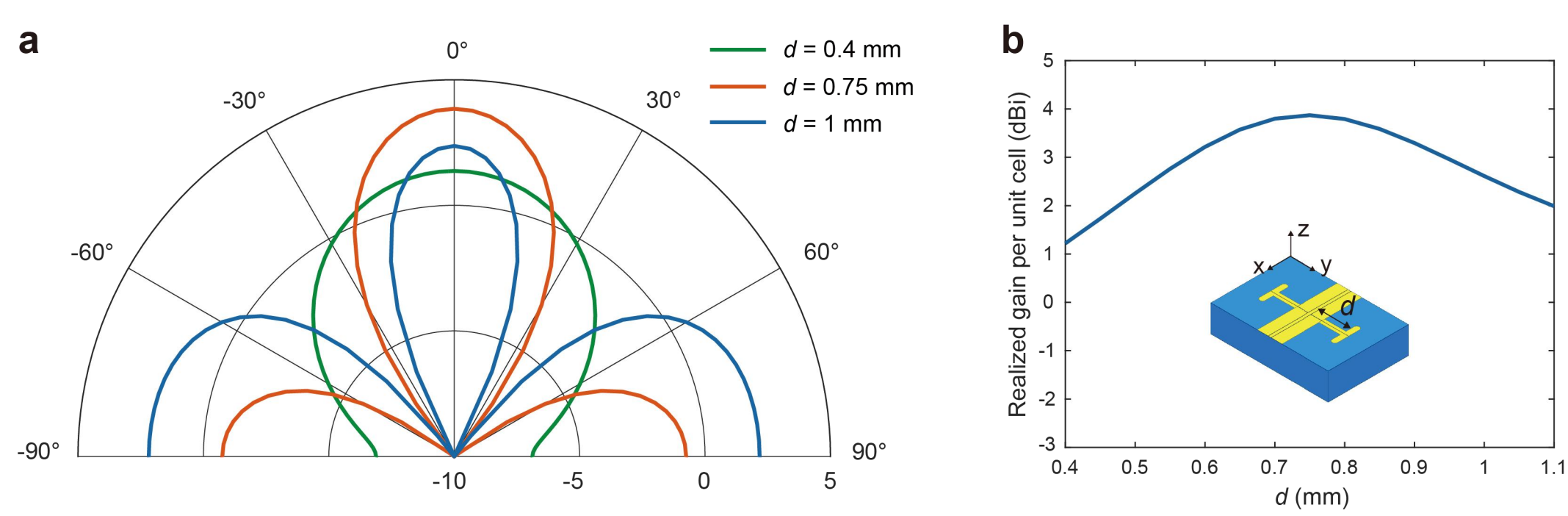

**Supplementary. Fig. S2 | Unit cell optimization. a,** Radiation patterns on the *yoz* plane for different *d*. **b.** Unit cell gain versus *d*, showing that there is an optimal value of *d*.

The unit cell comprises two mirror-arranged dipole antennas with a spacing *d* between the antenna and the central plane of the waveguide. The dipole element is fed by the slot etched on the ground electrode. **Supplementary Fig. S2a** investigates the radiation patterns with different spacings. As *d* increases from 0.4 mm to 0.75 mm, the beam is condensed, forming a narrow beam at the broadside with higher gain. The beamwidth continues to decrease as *d* further increases from 0.75 mm to 1 mm. However, the antenna gain decreases as *d* further increases because the grating lobes at the horizontal directions reduce the antenna directivity. As a result, there is an optimal value for *d*. **Supplementary Fig. S2b** simulates the broadside antenna gain at the D-band with different *d*. It shows that the optimal element spacing is achieved near *d* = 750 µm.

## Supplementary Note 4: Angle-robust broadband free-space THz-optic conversion

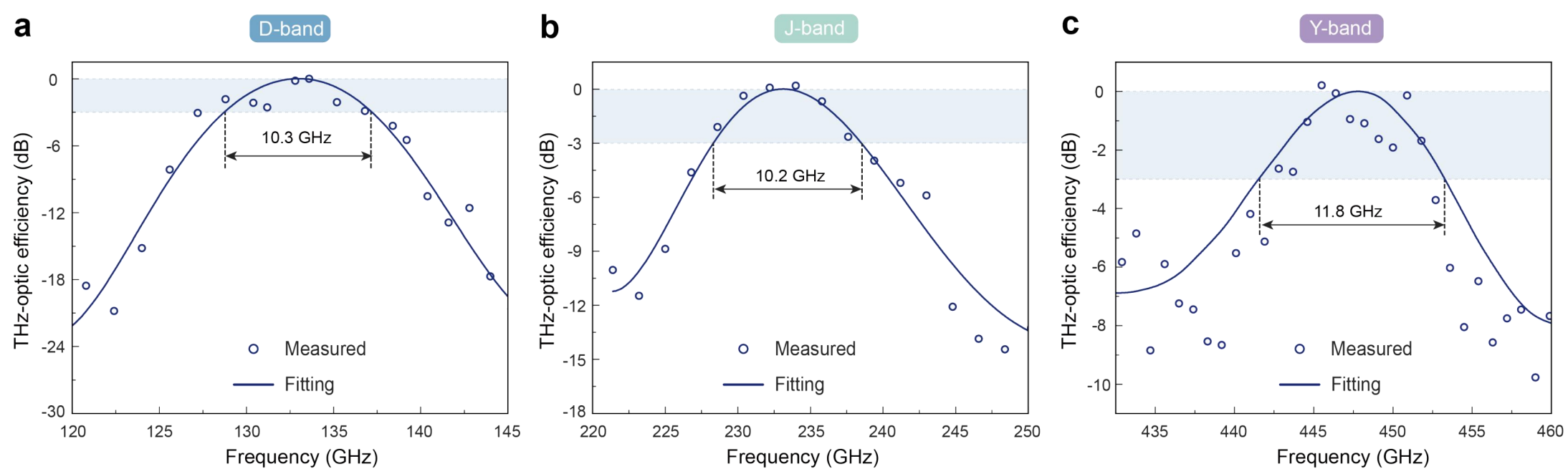

**Supplementary. Fig. S3 | Consistent THz-optic conversion bandwidths of the chips.** Normalized CSR versus incident THz frequency at the peak-CSR angle for the D-band (a), J-band (b) and Y-band (c) chips.

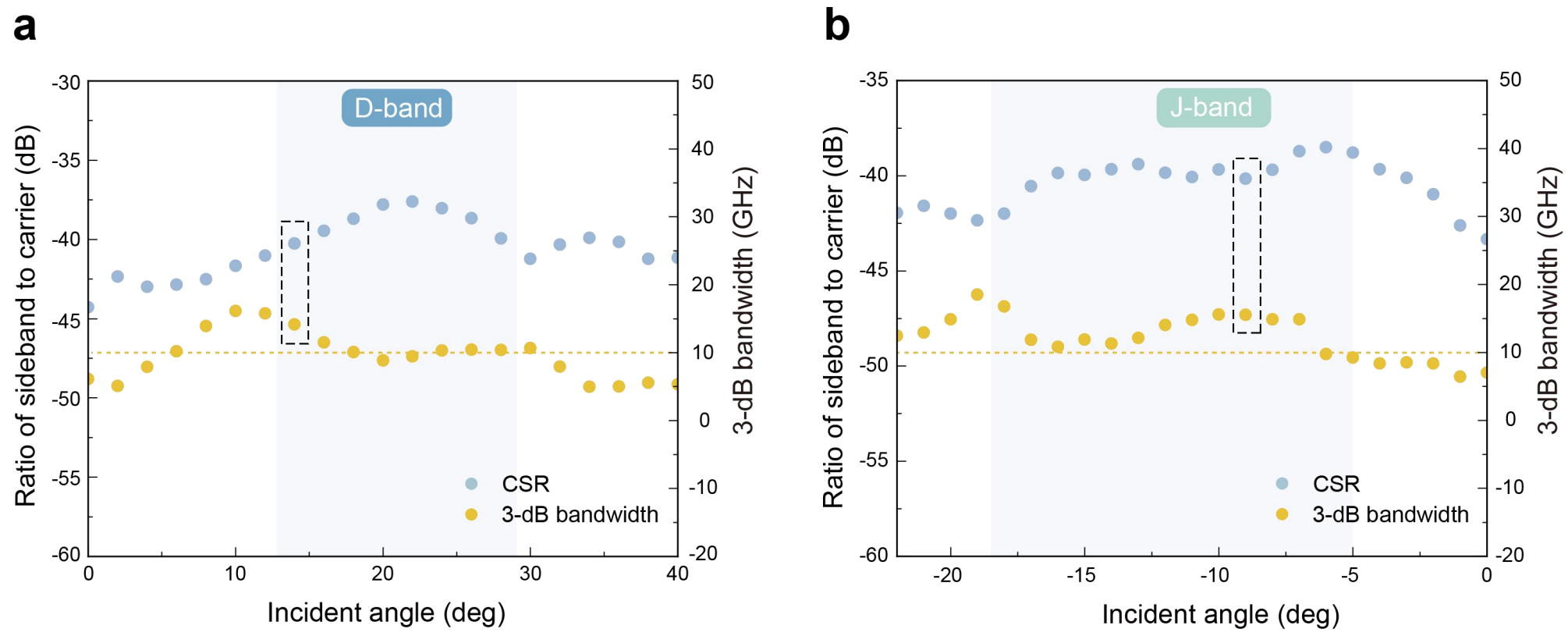

**Supplementary. Fig. S4 | Angle-robust broadband free-space THz-optic conversion. a-b,** CSR and 3-dB bandwidth versus incident angle for the D-band (a) and J-band chip (b).

The peak CSRs are achieved at 22°, -6° and -23° for the D-band, J-band and Y-band chips, respectively, all with a bandwidth of around 10 GHz, as shown in **Supplementary Fig. S3**. Moreover, taking 3 dB as the criteria, the D-band and Y-band chips achieve angle-robust THz-optic conversion within an angular range from 13° to 29° and from -16° to -5° (shaded areas in **Supplementary Fig. S4**), respectively, all with a bandwidth about 10 GHz. By shifting the incident angle to 14° for the D-band and -9° for the J-band, the bandwidth can be extended to 14 GHz and 15.5 GHz, respectively, while maintaining a high CSR of -40.26 dB and -40.16 dB, respectively, as indicated in the dashed box in **Supplementary Fig. S4**.

**Supplementary Note 5: Distributed-driven architecture for enhanced free-space THz-optic conversion**

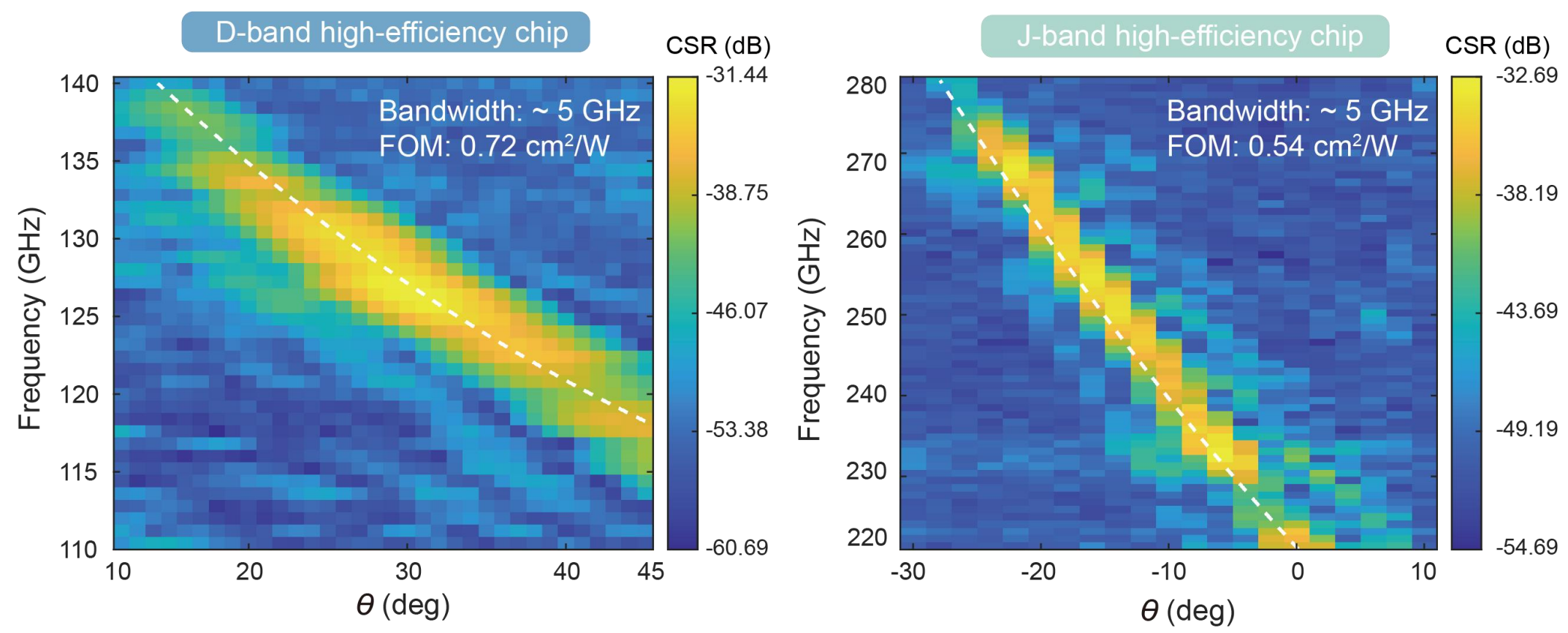

**Supplementary. Fig. S5 | Measurement results of high-efficiency chips.** Angle-frequency map of the measured CSRs for the D- and J-band high-efficiency chips.

As discussed in **Fig. 2**, the distributed-driven architecture makes the THz-optic conversion process less prone to the THz attenuation along the CPW electrode. As a result, the THz-optic conversion efficiency can be efficiently scaled up by increasing the effective interaction length. Benefit from the wafer-scale manufacturability in the TFLN platform, which provides an ideal toolbox to fabricate devices with different feature sizes, we extend the effective THz-optic interaction length of our D-band and J-band designs from 12 to 26 mm (corresponding to a theoretical 4.7× improvement) and measure their THz-optic conversion efficiency using the measurement setup in **Fig. 3a**. As shown in **Supplementary Fig. S5**, the measured results reveal a peak CSR of -31.44 and -32.69 dB when normalized to a free-space power density of 10 $W/m^2$. These numbers correspond to FOMs of 0.72 and 0.54 $cm^2/W$ for our D-band and J-band designs, respectively, showing both nearly 4× improvements while maintaining a ~5 GHz bandwidth. The achieved improvement is slightly lower than the theoretical value of 4.7×, probably due to the performance variation across different fabrication runs. Nonetheless, it has already represented over two orders of magnitude improvement over previous results (**Supplementary Table S1**) in the J- and Y-bands.

**Supplementary Note 6: Frequency-agnostic THz receiving link enabled by scalable THz-optic conversion**

According to the definition of EINF, the isotropic conversion gain is defined by:

$$EICG = \frac{P_{IF}}{P_r}$$

where $P_{IF}$ and $P_r$ are the PD IF output power and the received free-space power of an isotropic antenna. Assuming a unity input free-space power density (defined under unit input power $P_{in}$), the achieved IF power is proportional to:

$$\propto RP_c * FOM * P_{LO}L$$

where $R$, $P_c$, $P_{LO}$ and $L$ represent the responsivity of the photodetector, carrier power, LO power and optical loss factor. Note that the THz-optic conversion FOM is defined (see main text) regardless of the frequency band. It means the scalable FOMs result in consistent $P_{IF}$ across different bands, assuming fixed $P_{in}$, $R$, $P_c$, $P_{LO}$ and $L$.

On the other hand, the received THz power can be obtained by multiplying the free-space power density with effective antenna receive aperture $S = \frac{G\lambda^2}{4\pi}$, yielding:

$$P_r = \frac{P_{in} G \lambda^2}{4\pi}$$

indicating that $P_r$ is inversely scaled with increasing frequency (reducing $\lambda$) since $G$ is always equal to unity for an isotropic antenna. The consistent $P_{IF}$ and reduced $P_r$ together result in an increasing conversion gain with frequency, revealing a positively-scaled link performance. Our measured conversion gain confirms this prediction, as shown in **Supplementary Fig. S6**.

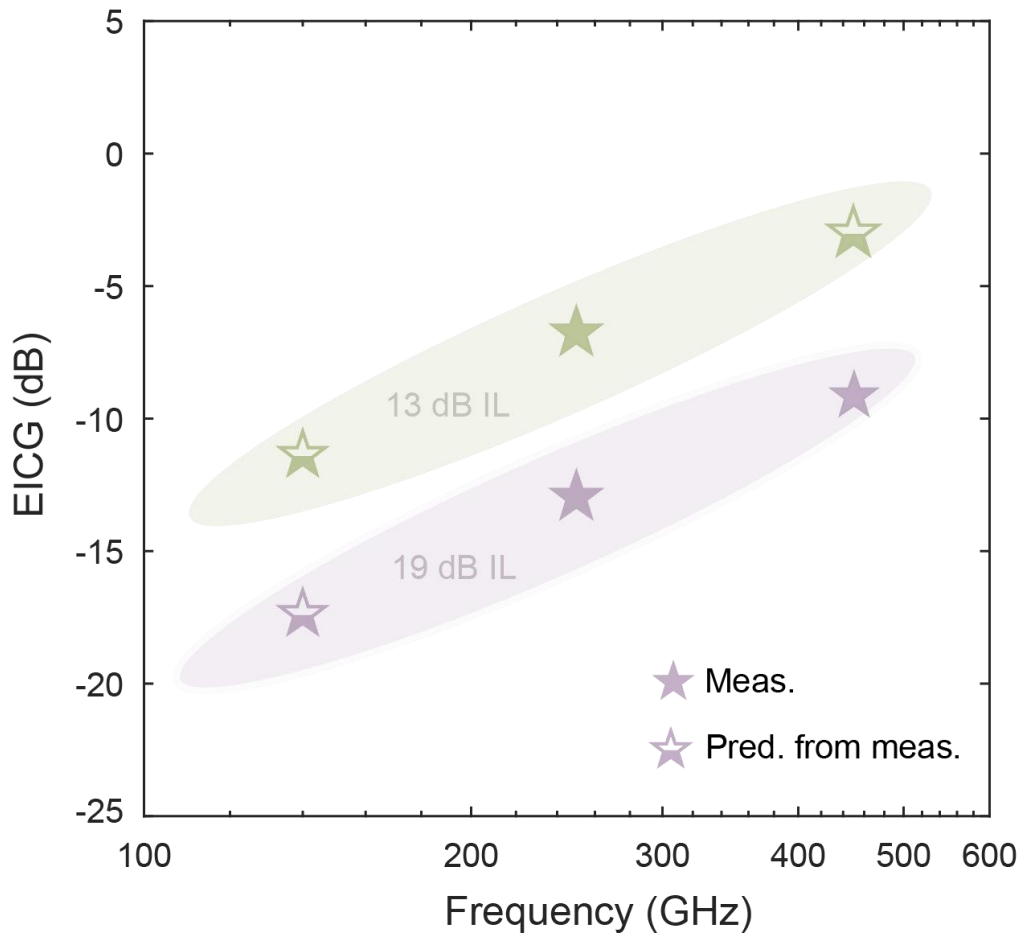

**Supplementary. Fig. S6 | Effective isotropic conversion gain at different bands.**

## Supplementary Note 7: Broadband down-conversion in the Y-band

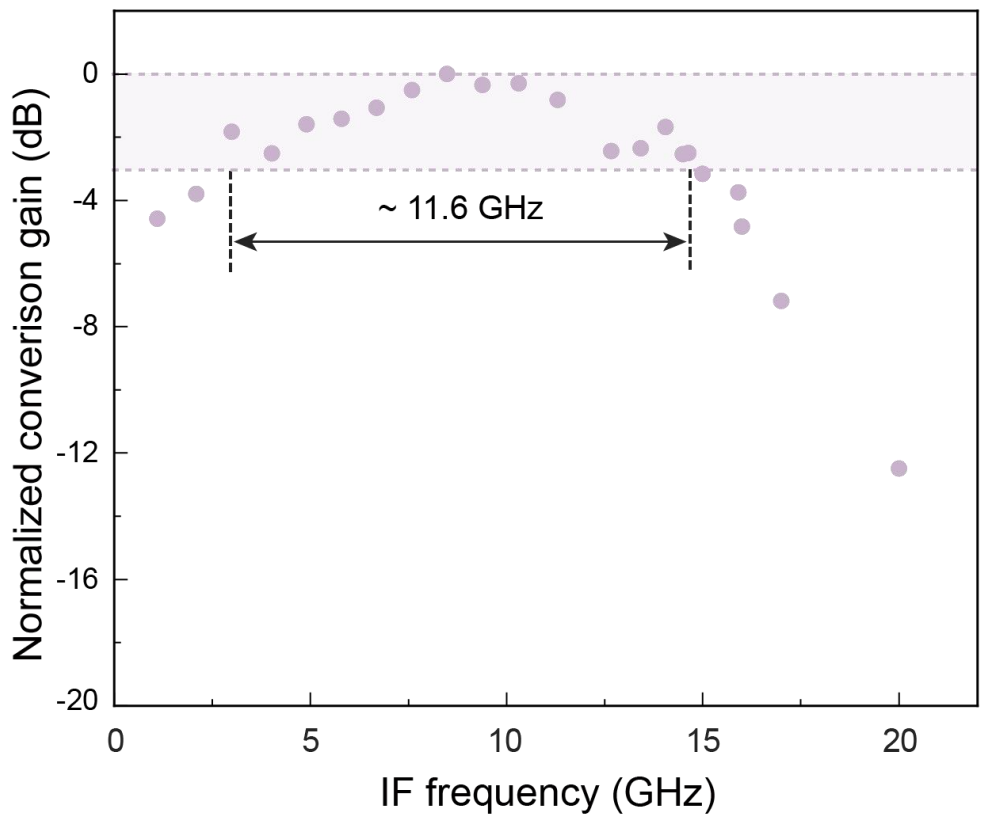

**Supplementary. Fig. S7 | Broadband down-conversion at 450 GHz.** Measured conversion gain bandwidth with a fixed LO spaced by 440 GHz from the carrier.

**Supplementary Fig. S7** shows the measured link RF conversion gain bandwidth with a fixed LO laser spaced by 440 GHz from the carrier while sweeping RF frequencies. With a 11.6 GHz 3-dB link conversion bandwidth and superior EINF, our Y-band frontend effectively outperforms state-of-the-art-electronics in both key metrics.

**Supplementary Note 8: Comparison to state-of-the-art electronics**

**Supplementary Tables S3** compares the link performance of full receiving frontends. **Supplementary Fig. S8** further compares the equivalent antenna-less link noise figure performance to state-of-the-art electronics (**Supplementary Table S4**). Our integrated frontends outperform state-of-the-art integrated electronics in both J- and Y-bands. By employing double-layer tapered couplers or heterogeneous integration techniques to eliminate the external coupling loss and assume a 30 dBm pump power, our integrated frontend could achieve ultralow EINFs of -6.2 and -9.8 dB in J- and Y-bands, respectively, a regime inaccessible to current integrated electronic solutions. The corresponding NFs are 9.1 dB, 8.1 dB, and 3.2 dB at 140, 250 and 450 GHz, significantly surpassing state of the art.

**Supplementary Table S3 Performance comparison with the full receiving frontends**

| Ref. | Platform | Operating frequency (GHz) | EINF (dB) | Conversion gain bandwidth (GHz) |
|---|---|---|---|---|
| [16] | CMOS | 260 | 15.8 | N/A |
| [S1] | BiCMOS | 335 | 19.7 | N/A |
| [15] | CMOS | 380 | 17.7 | N/A |
| [S2] | CMOS | 402 | 31.5 | 0.8 |
| [S3] | CMOS | 410 | 29.7 | N/A |
| [S4] | CMOS | 425 | 47.7 | 1.2 |
| [S5] | CMOS | 431 | 30.5 | 1.1 |
| [S6] | CMOS | 460 | 17.2 | 6.5 |
| [S7] | CMOS | 490 | 42.7* | N/A |
| [S8] | CMOS | 490 | 51.1 | 0.017 |
| [S9] | CMOS | 560 | 35 | N/A |
| [S10] | CMOS | 607 | 64 | 1 |
| **This work** | **TFLN photonics** | **250** | **13.6 / -6.2†** | **10.2** |
| | | **450** | **16.2 / -9.8†** | **11.6** |

* Calculated from the reported minimum discernible signal

† Predicted from measured results with high-efficiency chips, 2 dB optical IL and 30 dBm pump power

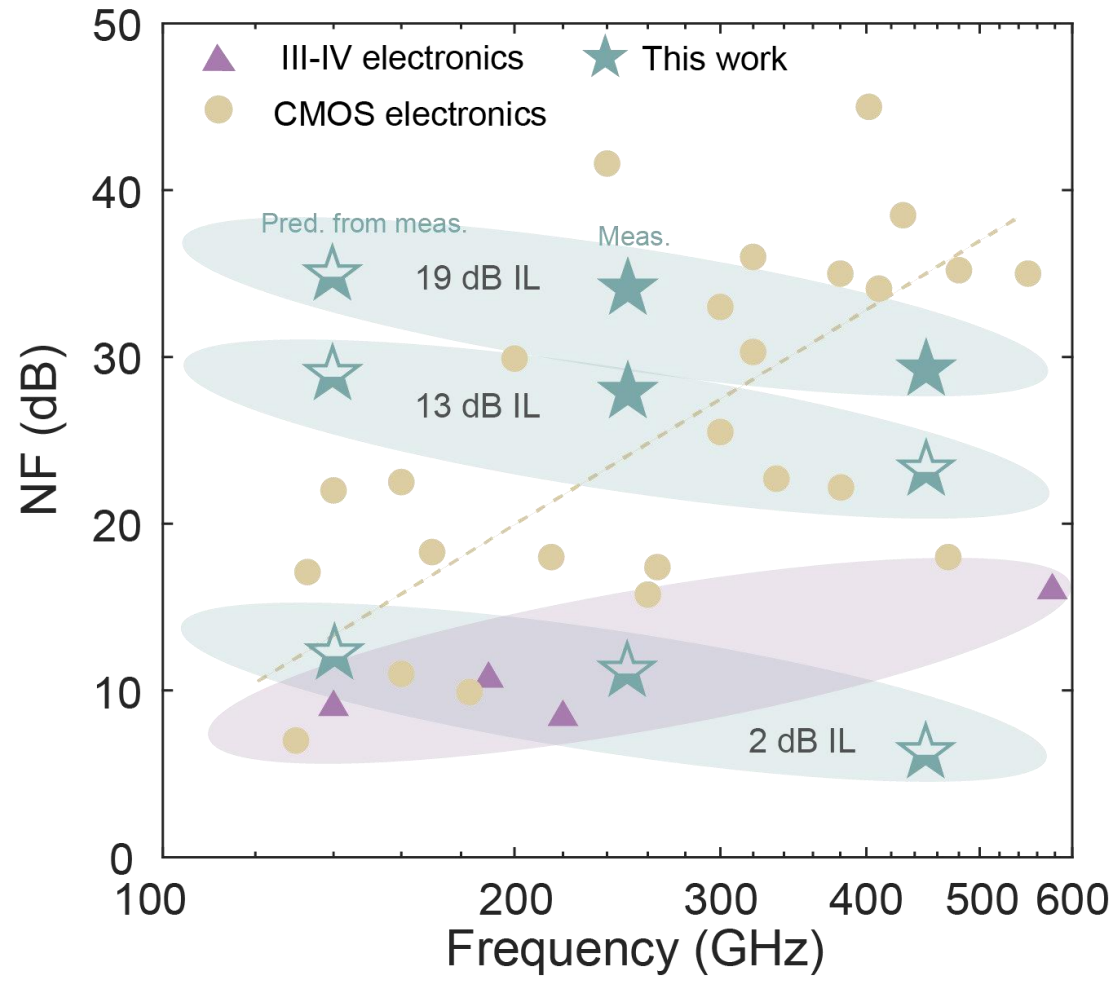

**Supplementary. Fig. S8 | Comparison of noise figure to the state-of-the-art electronics.**

**Supplementary Table S4 Comparison of the detection link NF**

| Title | Platform | Operating frequency (GHz) | NF (dB) |
|---|---|---|---|
| A D-Band Front-End T/R MMIC in a 70-nm GaN HEMT Technology | III V | 130 | 7 |
| A D-Band mm-wave Spectroscopy TX and RX in 28 nm CMOS with 15.6 dBm EIRP and 17.1 dB NF with Integrated Antennas | CMOS | 133 | 17.1 |
| A 1.2V, 140GHz Receiver with On-Die Antenna in 65nm CMOS | CMOS | 140 | 22 |
| A Highly Integrated Chipset for 40 Gbps Wireless D-band Communication Based on a 250 nm InP DHBT Technology | III V | 140 | 9 |
| A 160 GHz Pulsed Radar Transceiver in 65 nm CMOS | CMOS | 160 | 22.5 |
| A SiGe Quadrature Transmitter and Receiver Chipset for Emerging High-frequency Applications at 160 GHz | CMOS | 160 | 11 |
| 130-320-GHz CMOS Harmonic Down converters around and above the Cutoff Frequency | CMOS | 170 | 18.3 |
| 183GHz 13.5mW/Pixel CMOS Regenerative Receiver for mm-Wave Imaging Applications | CMOS | 183 | 9.9 |
| A low-power SiGe BiCMOS 190-GHz Transceiver Chipset with Demonstrated Data Rates Up to 50 Gbit/s Using On-chip | CMOS | 190 | 10.7 |
| antennas | | | |
| A 200 GHz Downconverter in 90 nm CMOS | CMOS | 200 | 29.9 |
| Subharmonic 220- and 320-GHz SiGe HBT Receiver Front-Ends | CMOS | 215 | 18 |
| A 220 GHz Single-Chip Receiver MMIC with Integrated Antenna | III V | 220 | 8.4 |
| A 32-unit 240-GHz Heterodyne Receiver Array in 65-nm CMOS with Array-wide Phase Locking | CMOS | 240 | 41.6 |
| A 220-to-320-GHz FMCW Radar in 65-nm CMOS Using a Frequency-Comb Architecture | CMOS | 260 | 15.8 |
| A 76-Gbit/s 265-GHz CMOS Receiver With WR-3.4 Waveguide Interface | CMOS | 265 | 17.4 |
| 32-Gbit/s CMOS Receivers in 300-GHz Band | CMOS | 300 | 25.5 |
| 300-GHz CMOS Receiver Module with WR3.4 Waveguide Interface | CMOS | 300 | 33 |
| A 0.32 THz FMCW Radar System Based on Low-Cost Lens-Integrated SiGe HBT Front-Ends | CMOS | 320 | 30.3 |
| Subharmonic 220- and 320-GHz SiGe HBT Receiver Front-Ends | CMOS | 320 | 36 |
| Fully Integrated Single-Chip 305–375-GHz Transceiver with On-Chip Antennas in SiGe BiCMOS | CMOS | 335 | 22.7 |
| A 0.38 THz fully integrated transceiver utilizing a quadrature push-push harmonic circuitry in SiGe BiCMOS | CMOS | 380 | 35 |
| THz prism: One-shot simultaneous localization of multiple wireless nodes with leaky-wave THz antennas and transceivers in CMOS | CMOS | 380 | 22.2 |
| A 400-GHz 28-nm TX and RX with Chip-to-Waveguide Transitions Used in Fully Integrated Lensless Imaging System | CMOS | 402 | 45 |
| 410 GHz CMOS Imager Using a 4th Sub-Harmonic Mixer with Effective NEP of 0.3 $fW/Hz^{0.5}$ at 1-kHz Noise Bandwidth | CMOS | 410 | 34.1 |
| 426-GHz Concurrent Transceiver Pixel in 65-nm CMOS for Active Imaging | CMOS | 425 | 54.3 |
| A 430 GHz CMOS Concurrent Transceiver Pixel Array for High Angular | CMOS | 430 | 38.5 |
| Resolution Reflection-mode Active Imaging | | | |
| A 460-GHz Receiver Using Second-Order Subharmonic Mixer in 65-nm CMOS | CMOS | 460 | 18 |
| A Fully Integrated 0.48 THz FMCW Radar Sensor in a SiGe Technology | CMOS | 480 | 35.2 |
| A fully Integrated 490-GHz CMOS Receiver Adopting Dual-locking Receiver Based FLL | CMOS | 490 | 58.1 |
| Terahertz RF Front-End Employing Even-Order Subharmonic MOS Symmetric Varactor Mixers in 65-nm CMOS for Hydration Measurements at 560 GHz | CMOS | 560 | 35 |
| InP HBT Technologies for THz Integrated Circuits | III V | 577 | 16 |
| A 4×4 607 GHz Harmonic Injection Locked Receiver Array Achieving 4.4pW/√Hz NEP in 28nm CMOS | CMOS | 607 | 72 |
| A 0.68-THz receiver with third-order sub-harmonic mixing in 65-nm CMOS | III V | 680 | 28.4 |
| **This work** | **TFLN photonics** | **140**<br>**250**<br>**450** | **N.A. / 12.1† / 9.1‡**<br>**27.9 / 11.1† / 8.1‡**<br>**29.2 / 6.2† / 3.2‡** |

† Predicted from measured results with high-efficiency chips, 2 dB optical IL

‡Predicted from measured results with high-efficiency chips, 2 dB optical IL and 30 dBm pump power